On the origin of the E1 electron trap level in GaN and dilute $Al_xGa_{1-x}N$ films

P. Kruszewski[1], J. Coutinho[2], V.P. Markevich[3], P. Prystawko[1], L. Sun[3], J. Plesiewicz[1],

C.A. Dawe[3], M.P. Halsall[3], and A.R. Peaker[3]

[1] Institute of High Pressure Physics, Polish Academy of Sciences,
Sokolowska 29/37, 01-142 Warsaw, Poland

[2] i3N, Department of Physics, University of Aveiro, Campus Santiago,
Aveiro 3810-193, Portugal

[3] Photon Science Institute and Department of Electrical and Electronic Engineering,
the University of Manchester, Manchester, M13 9PL, UK

Corresponding author: kruszew@unipress.waw.pl

**Abstract**

The results of high-resolution Laplace deep-level transient spectroscopy (L-DLTS) measurements applied to the E1 and E3 {$Fe_{Ga}(0/-)$} electron traps in dilute $Al_xGa_{1-x}N$ films (x = 0.063), grown by metal-organic vapor phase epitaxy (MOVPE) on Ammono-GaN substrates, are presented. It is shown that the electron emission signals associated with the E1 donor and the $Fe_{Ga}(0/-)$ acceptor levels split into individual components due to the aluminium fluctuations in the nearest neighbour shells around the E1 and $Fe_{Ga}$ defects. The splitting patterns observed in the L-DLTS spectra are nearly identical for both signals. Furthermore, the ratios of peak magnitudes determined from the L-DLTS analysis for both the E1 and E3 traps are consistent with calculated probabilities of finding a given number of aluminium atoms in the second nearest neighbour shell around a Ga lattice site in $Al_xGa_{1-x}N$ with x = 0.063. These findings provide strong evidence that both the E1 and the $Fe_{Ga}(0/-)$ trap states in dilute $Al_xGa_{1-x}N$ are related to defects located in the Ga sublattice. To elucidate the origin of the E1 trap in $Al_xGa_{1-x}N$, we have performed a comprehensive scan of possible impurities and defects in GaN and $Al_xGa_{1-x}N$ using hybrid density functional calculations of transition levels and their associated shifts upon substitution of Ga neighbour atoms by Al. From analysis of the results, we find that the E1 electron trap in GaN and $Al_xGa_{1-x}N$ is most likely related to a donor transition from a carbon or molybdenum impurity atom at the gallium site, $C_{Ga}(0/+)$ or $Mo_{Ga}(0/+)$, respectively.

It is well documented that defects or impurities can have a detrimental impact on the properties of gallium nitride (GaN) based materials. This, in turn, may result in lower efficiency and reliability of electronic and optoelectronic devices, such as $Al_xGa_{1-x}N$/GaN high electron mobility transistors (HEMTs),[1,2] lasers[3,4] and power diodes.[5,6]

The most prominent electron traps commonly detected by DLTS (Deep Level Transient Spectroscopy) and Laplace-DLTS (L-DLTS)[7] in n-type GaN are the so-called E1 and E3 emission signals.[8-13] The E3 trap in n-type GaN films grown by metal-organic vapor phase epitaxy (MOVPE) has been recently assigned to the $Fe_{Ga}(0/-)$ acceptor level,[14] while the origin of E1 remains inconclusive. The previously proposed assignments for E1 primarily rely on studies of n-GaN films grown on lattice-mismatched sapphire substrates.[10,15-18] The DLTS and L-DLTS emission signals in such structures are significantly broadened, making the interpretation of the results ambiguous. Fang et al.[15] studied highly dislocated HVPE GaN films ($\sim 10^9$-$10^{10}$ cm$^{-2}$) grown on $Al_2O_3$ and found a correlation between the density of dislocations with the concentration of gallium vacancies ($V_{Ga}$). Consequently, they suggested that an electron trap labelled as D, with trap signatures close to those of E1, could be a complex involving $V_{Ga}$, such as $V_{Ga}$–$V_N$.[15] This suggestion implies that in n-GaN films with low dislocation density ($\sim 10^5$ cm$^{-2}$), the concentration of the E1 should be low, which contradicts various recent observations.[11,13,19-22] In turn, other groups that studied molecular beam epitaxy (MBE) grown GaN films have attributed the E1 trap to a $V_N$-related defect.[16,20,23] However, this attribution is inconsistent with the results of recent studies, showing that homoepitaxially grown GaN films, subjected to electron or gamma-ray irradiation, did not exhibit significant changes in magnitude of the E1 signal.[11,20] The results of the irradiation experiments indicate that it is unlikely that the E1 trap is related to native point defects such as vacancies or interstitials. More detailed discussion of the origin of the E1 can be found in Ref. 17. A recent study found that the electron emission from the E1 trap in MOVPE grown n-GaN layers on native substrates is substantially enhanced by an applied electric field, which is characteristic for a donor-type level.[13] Various aspects of defects in GaN can be found in recent reviews.[24,25]

There is little information on the behaviour of defect energy levels upon changes of Al content in $Al_xGa_{1-x}N$ materials. Most of the earlier studies have been carried out on $Al_xGa_{1-x}N$ films grown on sapphire, which complicated the analysis of alloying induced effects.[26-28] Fortunately, recent advancements have been made with $Al_xGa_{1-x}N$ films grown on native GaN substrates, such as Ammono-GaN.[29,30] Kruszewski et al.[31] reported an alloying effect in dilute $Al_xGa_{1-x}N$ films (with x ≤ 0.05), which is characterized by the splitting of the E3 emission signal in the L-DLTS spectra. This splitting arises from fluctuations in the number of aluminium (Al) atoms in the nearest neighbour shells surrounding the E3 defect. This finding is important as it highlights the capabilities of L-DLTS for precise investigations of the local environment around defects in $Al_xGa_{1-x}N$,[31] and potentially in other ternary systems as well. Furthermore, the peak magnitudes ratios determined from the L-DLTS analysis for the E3 traps in $Al_xGa_{1-x}N$ closely align with the calculated probabilities of finding a specific number of aluminium atoms in the second-nearest neighbour shell around a Ga lattice site for the given x values. These findings have provided compelling evidence that the E3 defect and its associated electronic level in $Al_xGa_{1-x}N$ are most likely related to a Fe atom substituting gallium,[31] thereby supporting the assignment of E3 to the $Fe_{Ga}(0/-)$ level in GaN, as proposed by Horita et al.[14]

Below, we will demonstrate that a DLTS signal associated with a level $E_C$–0.34 eV, which is assigned to the E1 trap in $Al_xGa_{1-x}N$ (x = 0.063),[32] is related to a point defect localized in the Ga sublattice.

In this study, Si-doped $Al_{0.063}Ga_{0.937}N$ film with a thickness of 1.5 μm was grown by the MOVPE technique on top of GaN:Si buffer layers deposited on a highly conductive Ammono-GaN substrate.[29,30] More details of the epi growth and contacts fabrication can be found in Refs. 13 and 19.

First-principles modelling of defects was performed within density functional theory (DFT), using a hybrid treatment of the exchange correlation potential as proposed by Heyd-Scuseria-Ernzerhof (HSE).[33] We followed the recipe proposed by Deák et al.[34] specially tested for the description of carbon-related centers in GaN. Accordingly, a fraction of exact exchange, $\alpha = 0.26$, is mixed with the semi-local potential, and that is phased out with a screening length $\mu = 0.1$ Å$^{-1}$. The calculations were executed using the Vienna Ab-initio Simulation Package (VASP).[35,36] The code employs the projector augmented-wave method for the description of core electrons, whereas valence-related scalar fields are expanded with plane waves and solved using the Kohn-Sham algorithm. Semicore Ga 3d states were part of the valence states, the plane wave cut-off for the wave functions and electron density was 400 eV and 800 eV, respectively. The GaN and $Al_xGa_{1-x}N$ hosts were idealized as hexagonal supercells with 300 atoms, which were found to be large enough to justify a gamma sampling of the Brillouin zone. See Ref.34 for further details.

The DLTS scans carried out on $Al_xGa_{1-x}N$ samples for temperatures between 80 K and 415 K revealed two trap levels: $E_c - 0.34$ eV and $E_c - 0.68$ eV, with concentrations of about $5\times10^{13}$ cm$^{-3}$ and $2.5\times10^{14}$ cm$^{-3}$, respectively (see supplementary materials). The electronic signatures of both defects correspond to the values for the E1 and E3 traps, which can be extrapolated to $Al_xGa_{1-x}N$ with x = 0.063, from the results reported by Sun et al.[32] and Kruszewski et al.[31] for $Al_xGa_{1-x}N$ with x≤ 0.05. Therefore, we assign the former trap level to the E1 trap while the latter to the $Fe_{Ga}(0/-)$ acceptor.[31]

Two L-DLTS spectra showing electron emission signals from the $Fe_{Ga}(0/-)$ acceptor (red line) and E1 donor (black line) states in $Al_xGa_{1-x}N$ with x = 0.063 are shown in Fig.1a). Both spectra were measured with identical experimental conditions specified in Fig.1a), and to satisfy a condition of nearly uniform electric field in the probed region, the Double L-DLTS technique was used.[7,37] Hence, the L-DLTS spectra are free from broadening related to the electric field inhomogeneity. As observed in Fig.1a), the L-DLTS spectrum for the $Fe_{Ga}(0/-)$ consists of three well-resolved peaks corresponding to 0, 1 and 2 Al atoms in the 2NN shell around the Fe atom, as reported previously.[31] The E1 spectrum exhibits four components.

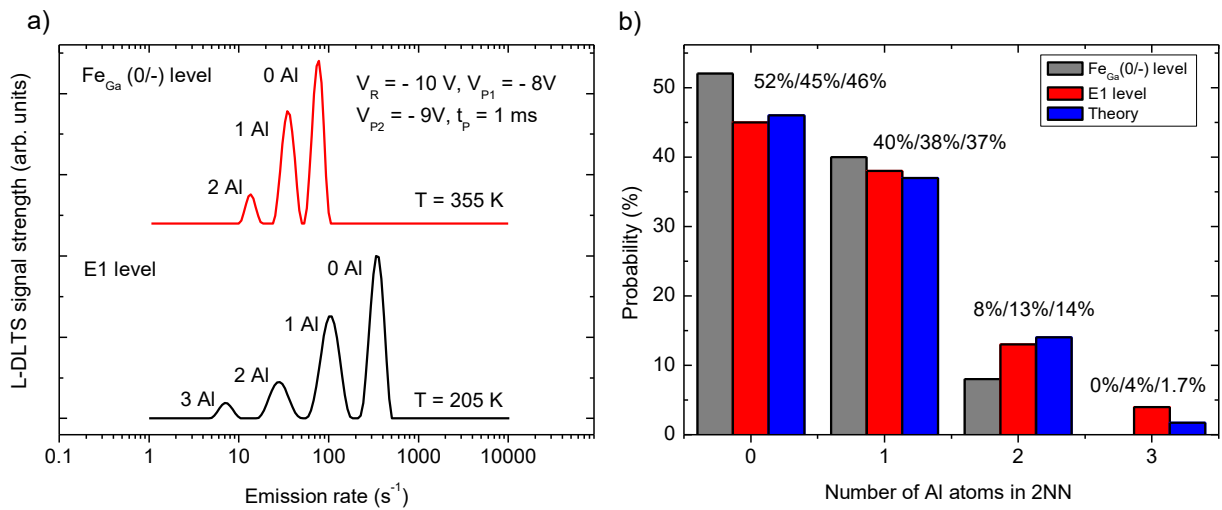

**Fig.1** a) The L-DLTS spectra for the $Fe_{Ga}(0/-)$ (red) and E1 trap level (black) measured at 355 and 205 K, respectively; b) The probabilities of finding a given number of Al atoms in the 2NN around the $Fe_{Ga}(0/-)$ acceptor (grey bars), and E1 trap level (red bars), calculated from the L-DLTS peaks magnitudes for $Al_xGa_{1-x}N$ with x = 0.063. The blue bars correspond to calculated probabilities assuming a fully random distribution of Al atoms in the crystal.

If one assumes that the E1 trap state, as in the case of $Fe_{Ga}(0/-)$, is related to a defect located in the Ga sublattice, then the four emission components observed in the L-DLTS spectrum could correspond to configurations with 0, 1, 2, and 3 Al atoms in the 2NN shell around the E1 defect. To validate this, we calculated and compared the theoretical and experimental probabilities of finding a given number of Al atoms in the 2NN shell of the E1 and $Fe_{Ga}$ for the $Al_{0.063}Ga_{0.937}N$ alloy having random distribution of Al atoms on Ga sites. Both, the theoretical and experimental probabilities are visualized as coloured bars in Fig.1b). It is worth mentioning that the experimental values were determined from the L-DLTS peak intensities as described in Ref.31. As shown in Fig.1b), the experimentally determined probabilities for the E1 defect (red bars) and the $Fe_{Ga}(0/-)$ level (grey bars) are fairly similar. The partial contribution values determined for the E1 trap are in excellent agreement with the theoretically calculated probabilities denoted as blue bars in Fig.1b). The values for 0, 1 and 2 Al atoms in the 2NN shell are almost identical for both the experimental and theoretical data, whereas the probability corresponding to 3 Al atoms seems to be slightly overestimated (4%). This overestimation could be explained by the low intensity of the L-DLTS signal component associated with three Al atoms in the 2NN shell, resulting in a higher uncertainty in this particular case.

It is important to note that the energy resolution of the L-DLTS technique is almost inversely proportional to the temperature at which the spectrum is measured.[7] Therefore, the E1 spectrum (measured at 205 K) was obtained with a factor of 1.7 higher emission-rate resolution compared to the $Fe_{Ga}(0/-)$ spectrum (measured at 355 K). Hence, the additional peak related to 3 Al atoms revealed in the E1 case is a result of much better experimental conditions.

The experimental data presented above for the E1 trap level aligns remarkably well with the theoretical predictions. We believe these results provide compelling evidence for the proposed origin of the E1 trap level in dilute $Al_xGa_{1-x}N$, suggesting that this level corresponds to a point defect located at the Ga site, and most likely possessing $C_{3v}$ symmetry. Our identification of the E1 signal in $Al_xGa_{1-x}N$ arising from a defect at the Ga site is highly significant, as it substantially narrows the range of potential candidates responsible for the E1 trap level.

To facilitate a direct comparison of the L-DLTS spectra for the E1 and E3 traps, the frequency scale from Fig.1a) has been converted into an energy-difference scale using the formula:

$$\Delta E = - k_B \times T \times \ln(e_n/e_{n0}), \qquad (1)$$

where $\Delta E$ is the energy with respect to that of the 0 Al peak for each center, $k_B$ is the Boltzmann constant, T is the temperature, $e_n$ is the emission rate in the original spectrum shown in Fig.1a), and $e_{n0}$ represents an arbitrary reference frequency (0 Al peak), here, $e_{n0} = 78$ s$^{-1}$ for the Fe$_{Ga}$(0/–) and 338 s$^{-1}$ for E1. This enables a direct comparison of the Al atoms influence on the E1 and Fe$_{Ga}$(0/–) defect energy levels, irrespective of variations in experimental conditions, such as the measurement temperature. This method has been widely reported in Refs. 31, 38-40.

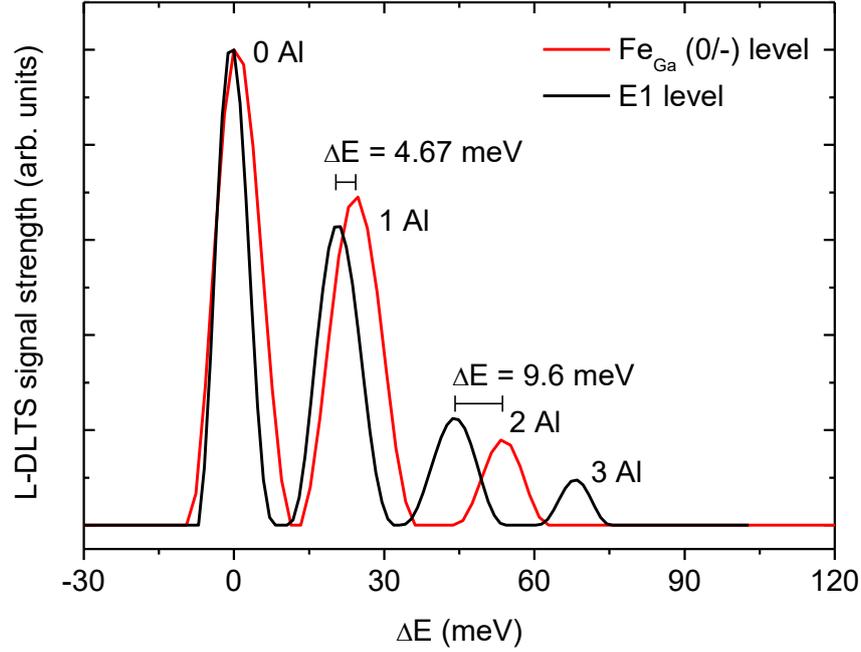

**Fig.2.** L-DLTS spectra for the Fe$_{Ga}$(0/–) (red) and E1 (black) in Al$_{0.063}$Ga$_{0.937}$N with the emission-rate scale recalculated to an energy-difference scale. The labels 0, 1, 2 and 3 Al in the figure correspond to 0, 1, 2 and 3 Al atoms in the 2NN shell centered around particular defect.

The normalized (to the emission signal of maximum magnitude) L-DLTS spectra recalculated to the energy scale using formula (1), are shown in Fig.2. Two observations can be drawn from this comparison: 1) the L-DLTS peaks corresponding to the electron emission from the Fe$_{Ga}$(0/–) defect exhibit a slight broadening when compared to the peaks in the spectrum for the E1 trap, and 2) the L-DLTS peaks for configurations with 1 and 2 Al atoms in the 2NN shell for Fe$_{Ga}$(0/–) are shifted to higher energies by 4.7 and 9.6 meV, respectively, relative to the corresponding E1 related peaks.

First, we suppose that the broadening of the L-DLTS peaks for the Fe$_{Ga}$(0/–) level is primarily due to much higher thermal effects; the $k_B \times T$ factor at 355 K is 30.6 meV, which is almost double the value of 17.7 meV at 205 K. This significant difference in thermal energy should contribute to the broadening of the L-DLTS signal.

As to the second observation, currently we propose two possible explanations: i) a different size of the substitutional impurities responsible for the Fe$_{Ga}$ and E1 defects could lead to variations in the elastic coupling with the Al neighbours. For each specific defect, a higher number of Al atoms in the 2NN shell would induce a larger elastic interaction. Alternatively, ii) the different shifts could be related to different electrostatic coupling between Al ions and the charged centers. Besides depending on the number of Al substitutions, such interactions could depend on the shape and extent of the donor/acceptor wavefunctions.

Modern first-principles methods can predict electronic transitions involving carrier traps with an accuracy of ~0.1 eV.[41] This hampers any attempt to calculate energy differences as low as ~10 meV (as determined above for the E1 and E3 traps with different numbers of Al atoms in the 2NN shell) without benefiting from systematic error cancelation. However, the experimentally determined positions of the E1 energy level in Al$_x$Ga$_{1-x}$N with $0 \leq x \leq 0.063$,[31,32] along with its identification as a donor transition (Ref.13) most likely from a highly symmetric center on the Ga site, and the observation that the transition deepens as Ga neighbors are substituted by Al atoms, provide key inputs for initiating the search for a possible defect model for E1.

Hybrid DFT calculations were performed for several chemical species with some probability of being incorporated in various GaN materials. These included Ga substitutions by B, C, N, O, Si, first-row transition metals from Ti to Zn, as well as heavier metals such as Zr, Mo and Hf. Numerous low-symmetry complexes were also explored, like carbon-dimers on Ga and N sites, oxygen-dimers and nitrogen-dimers on Ga, divacancy (V$_{Ga}$V$_N$), as well as nitrogen vacancy next to carbon on gallium site (V$_N$C$_{Ga}$).[42]

Among the studied defects, only C$_{Ga}$, Si$_{Ga}$, Ti$_{Ga}$, Mo$_{Ga}$ and V$_N$C$_{Ga}$ had donor levels in the upper half of the band gap. Other impurities either had donor transitions in the lower half of the gap or they were acceptors. Ti$_{Ga}$ shows a deep (0/+) transition at $E_c - 0.72$ eV, which is too deep to be considered a serious candidate for the E1 trap. In turn, Si$_{Ga}$ is a shallow donor, with a calculated level at 0.1 eV above the conduction band minimum.

The calculated donor transition of $V_N C_{Ga}$ is estimated at $E_c - 0.20$ eV, close to that for the E1 trap. The result agrees with recent reports by Matsubara and Bellotti.[42] We found that the (0/+) level of $V_N C_{Ga}$ with one Al$_{Ga}$ neighbor replacing one of the three Ga sites edging the vacancy is about 0.1 eV deeper compared to that of $V_N C_{Ga}$ with Al$_{Ga}$ at one of the other nine possible Al$_{Ga}$ lattice sites (closer to C). This follows from the anisotropy of the complex, which is not observed for the E1 trap.

In agreement with previous calculations,[43] we estimate $(0/+)$ and $(-/0)$ transitions for divacancy at $E_v - 1.21$ eV and $E_v - 1.61$ eV, respectively, allowing us to safely exclude $V_{Ga}V_N$ as a candidate for the E1 trap as well.

Finally, we are left with C$_{Ga}$ and Mo$_{Ga}$ centers with calculated donor transitions at $E_c - 0.34$ eV and $E_c - 0.20$ eV in GaN, respectively. Carbon is likely to be introduced from metal-organic precursors, whereas Mo could be a contaminant originating from the reactor vessel or from steel pipes. Both species are in principle good candidates to account for the E1 trap. Differences emerge only when we analyse alloying effects — Al replacement of a Ga nearest neighbor deepens the C$_{Ga}(0/+)$ transition by ~0.08 meV (in comparison to a similar supercell with Al at a remote location from C), while for the case of Mo$_{Ga}$ the $(0/+)$ transition shifts toward $E_c$ by 0.05 eV. While we note that the results for C$_{Ga}$ are in qualitative agreement with the results of L-DLTS measurements for the E1 trap, we cannot rule out a possible involvement of Mo based on such a subtle effect.

Previous calculations predicted a shallow donor character for C$_{Ga}$ in GaN.[42,44,45] In line with recent studies,[34,42] we found that explicit consideration of Ga 3d states in the valence states, leads to improvements in calculated energies and electrical levels by up to 0.3 eV. For the specific case of C$_{Ga}$ in GaN, the use of 96-atom supercells (as employed in Refs. 42 and 45) also resulted in a very shallow donor level for the C$_{Ga}$ impurity. However, after increasing the size of the cell to 300 atoms, we clearly find a deep state with a transition level at $E_c - 0.34$ eV. We attribute the above discrepancies to limitations imposed by the periodic boundary conditions, which are partially mitigated when we employ a larger supercell: (i) The larger volume improves the accommodation of strain induced by the rather short C-N bonds; (ii) A smaller Brillouin zone implies a lower dispersion of the donor state, which in the case of the smaller cells is overestimated.

In 2021, Karpov presented a thermodynamic model of carbon doping in n-GaN, based on reported first-principles calculated formation energies that were corrected to account for realistic MOVPE growth conditions.[46] Importantly, it was found that the Fermi level lies close to midgap and is not pinned to the $(-/0)$ level of the C$_N$ acceptors. Under these conditions, the formation energies (and thereby concentrations) of $C_N^-$ and $C_{Ga}^+$ become comparable, with a post-growth concentration of $C_{Ga}^+$ in the range of $10^{13}$-$10^{14}$ cm$^{-3}$ for total carbon concentration of about $5\times10^{16}$ cm$^{-3}$ in n-type GaN. This finding agrees well the E1 concentrations determined from the DLTS measurements in our GaN and Al$_x$Ga$_{1-x}$N samples grown by MOVPE.[32]

In summary, the results of the high-resolution L-DLTS measurements applied to both the E1 and Fe$_{Ga}$(0/−) electron traps in dilute Al$_x$Ga$_{1-x}$N films (x = 0.063) grown by the MOVPE technique on Ammono-GaN substrates are presented and discussed. We have shown that the theoretically calculated probabilities of finding a given number of Al atoms in the 2NN shell around a Ga lattice site are in excellent agreement with the experimental values determined from the L-DLTS peaks intensities for the E1 trap level. This finding provides strong evidence that the E1 center in dilute Al$_x$Ga$_{1-x}$N is a defect located in the Ga sublattice. This identification is particularly important, as it substantially narrows the range of potential candidates responsible for the E1 trap, not only in dilute Al$_x$Ga$_{1-x}$N but also in pure GaN layers. A comparative analysis of the experimentally obtained results with the results of our DFT calculations suggest that the E1 trap in Al$_x$Ga$_{1-x}$N materials is most probably related to a donor level of a single carbon atom at a Ga site, C$_{Ga}$(0/+).

## SUPPLEMENTARY MATERIALS

The electrical parameters of Al$_x$Ga$_{1-x}$N Schottky barrier diodes (SBDs), the procedure for DLTS measurements, and the electronic signatures determined for the E1 and E3 trap levels can be found in the supplementary materials.

## ACKNOWLEDGMENTS

P.K., J.P. and P.P. would like to thank the National Science Centre, Poland, for financial support through Project No. 2020/37/B/ST5/02593. J.C. acknowledge the FCT through projects Refs: LA/P/0037/2020, UIDB/50025/2020, UIDP/50025/2020, as well as the RNCA through Project No: 2024.04745.CPCA.A1 (Oblivion supercomputer). The authors in Manchester would like to acknowledge support by the EPSRC-UK under Contract Nos. EP/T025131/1 and EP/S024441/1. C.A.D. would also like to acknowledge support from the EPSRC CDT in Compound Semiconductor Manufacturing. The authors would like to thank Ewa Grzanka from the Institute of High Pressure Physics, PAS (Warsaw/Poland) for the estimation of the Al content in Al$_x$Ga$_{1-x}$N film.

**ORCID iDs**

P. Kruszewski: https://orcid.org/0000-0002-5301-9202

J. Coutinho: https://orcid.org/0000-0003-0280-366X

V.P. Markevich: https://orcid.org/0000-0002-2503-6144

P. Prystawko: https://orcid.org/0000-0003-0762-5639

L. Sun : https://orcid.org/0000-0003-1293-8445

J. Plesiewicz: https://orcid.org/0009-0001-4486-1422

C.A. Dawe: https://orcid.org/0009-0001-7860-1047

M.P. Halsall: https://orcid.org/0000-0001-7441-4247

A.R. Peaker: https://orcid.org/0000-0001-7667-4624

AUTHOR DECLARATIONS

**Conflict of Interest**

The authors have no conflicts to disclose.

DATA AVAILABILITY

The data that support the findings of this study are available from the corresponding author upon reasonable request.

AUTHOR CONTRIBUTIONS

**P. Kruszewski**: Conceptualization (lead); Data curation (lead); Formal analysis (lead); Funding acquisition (lead); Investigation (lead); Methodology (lead); Validation (lead); Visualization (lead); Writing – original draft (lead); Writing – review & editing (lead); **J. Coutinho**: Formal analysis; Investigation; Methodology; Validation; Visualization; Writing; **V.P. Markevich**: Formal analysis (equal); Investigation (equal); Validation (equal); Writing – review & editing (equal); **P. Prystawko**: Formal analysis (supporting); Resources (lead); **L.Sun**: Investigation (supporting); Writing – review & editing (supporting); **J. Plesiewicz**: Formal analysis (supporting); **C.A. Dawe**: Investigation (supporting); Writing – review & editing (supporting); **M.P. Halsall**: Formal analysis (supporting); Investigation (supporting); **A.R. Peaker**: Formal analysis (supporting); Supervision (supporting); Writing – review & editing (equal)